# Superrecursive Features of Interactive Computation

Mark Burgin

University of California, Los Angeles
Los Angeles, CA, 90095, USA

**Abstract**

Functioning and interaction of distributed devices and concurrent algorithms are analyzed in the context of the theory of algorithms. Our main concern here is how and under what conditions algorithmic interactive devices can be more powerful than the recursive models of computation, such as Turing machines. Realization of such a higher computing power makes these systems superrecursive. We find here five sources for superrecursiveness in interaction. In addition, we prove that when all of these sources are excluded, the algorithmic interactive system in question is able to perform only recursive computations. These results provide computer scientists with necessary and sufficient conditions for achieving superrecursiveness by algorithmic interactive devices.

## 1   Introduction

There is a tendency to oppose algorithms and interaction (cf., for example, [17]). This opposition is based on a very restricted understanding of algorithms, which is based on the Church-Turing Thesis that equates algorithms with Turing machines or other mathematical schemas that give rules for computation of a function. Some researchers claim that interactive computation is more powerful than Turing machines (cf., for example, [6, 7, 14, 15, 17]), while others insist that the Church-Turing Thesis still holds (cf., for example, [9]). However, contemporary understanding extends the concept of algorithm, making it closer to the general usage of the word "algorithm". Namely, algorithm is informally perceived as a (finite) structure (e.g., a system of rules) that contains for some performer (class of performers) exact information (e.g., instructions) that allows some performer(s) to pursue a definite goal (cf., for example, [3, 4]).

People need information that is contained in algorithms to make their activity efficient and purposeful. Consequently, one main achievement of $20^{th}$ century scientific thought was elaboration of the theory of algorithms and computation. This theory studies abstract and real automata, computers and networks, computation and communication. In many ways, this theory is the central cornerstone for computer science. Many key accomplishments in the theory of algorithms and computation converge to the famous Church-Turing Thesis, a statement determining the boundaries of algorithmic computations. The Church-Turing Thesis has long been considered as the most fundamental law within computing. However, recent developments in the theory of algorithms allow overcoming limitations in the Church-Turing Thesis. New mathematical models for algorithms and computation have appeared that extend prior theory in a manner similar to the way relativity theory and quantum mechanics went beyond Newtonian mechanics. These new models are more powerful than the classical recursive algorithm models, i.e., Turing machines, partial recursive functions, Lambda-calculus, and cellular automata.

*Algorithms and automata that are more powerful than Turing machines are called super-recursive.*

*Computations that cannot be realized or simulated by Turing machines are called hyper-computations.*

At the first glance, it looks like interactive systems are essentially different from computers and cannot be represented by computing models, such as Turing machines. Really, as Leeuwen and Wiedermann write [13], the purpose of an interactive system is usually not to compute some final result but to react to or interact with the environment in which the system is placed and to maintain a well-defined action-reaction behavior. Interactive systems are always operating and thus, may be seen as machines on infinite strings, but differ in the sense that their inputs are not specified and may depend on intermediate outputs and external sources.

However, if we consider only systems that work with symbolic information, then reaction to or interaction with the environment or with another such system is information transformation and exchange, or communication. In other words, functioning of and interactive system that works with symbolic information consists of computation and communication processes.

In this paper, we analyze sources of an interactive recursive algorithm/machine ability to outperform (have more computing power than) conventional Turing machines, that is, to be able to compute recursively non-computable functions. We find five such sources of interactive superrecursiveness. Namely, interactive superrecursiveness is possible when: 1) the interactive algorithm is itself super-recursive; 2) (Proposition 1) the interactive algorithm is recursive but contains initial information about some recursively non-computable function (has a non-recursive oracle); 3) (Proposition 2) the interactive recursive algorithm interacts with a super-recursive algorithm (a non-recursive environment); 4) (Theorems 2, 3 and 4) time of interaction is not recursively coordinated; 5) (Theorems 5) communication space is not recursively coordinated. The first three cases are not interesting because either superrecursive power comes not from interaction (case 1) or as it is well known, if a recursive device have access to a super-recursive information, then this device can compute recursively non-computable functions.

However, after finding sources of interactive superrecursiveness, it is natural to ask a question if all sources have been found or there are other sources that we have not been able to see. To show that our analysis of interactive superrecursiveness is complete, we prove (Theorems 6) that if interacting algorithms/devices are recursive and their interaction is organized/controlled by a recursive device/algorithms, then computable functions are also recursive.

Thus, we consider algorithms in the form of rules and devices that perform simple and constructive operations at each step and give a result after a finite number of steps (in finite time).

All algorithms are divided into three big classes [3]: *subrecursive*, *recursive*, and *super-recursive*. Algorithms and automata that have the same computing/accepting power [4] as Turing machines are called *recursive*. Examples are partial recursive functions or random access machines.

Algorithms and automata that are weaker than Turing machines, i.e., that can compute fewer functions, are called *subrecursive*. Examples are finite automata, context free grammars or push-down automata.

Algorithms and automata that are more powerful than Turing machines are called *super-recursive*. Examples are inductive Turing machines, Turing machines with oracles or finite-dimensional machines over the field of real numbers.

It is evident that if an interacting algorithm is super-recursive, then even without any interaction, it can perform hypercomputation. Thus, the main question is when taking a recursive algorithm (automaton) and providing for it means for interaction, we can achieve hypercomputation as a result of interaction. There are different models of interactive computational systems: persistent Turing machines [7], a global Turing machine or Internet machine [13, 14], Web machines [1], Web automata [11] and others. The most general, flexible and powerful model of interactive (computational) systems is a grid automaton [2, 3].

In the analysis of the computational power of interaction, it is possible to consider only two systems as interaction of any finite number of systems can be reduced by induction to the case of two systems. Here we do not consider infinite systems.

By the definition of a recursive algorithm (device or machine), a Turing machine can compute any function that a recursive algorithm can compute. That is why we can use Turing machines to explore problems of computational power of interaction. In addition, it is possible to consider Turing machines with one working tape as Turing machines with *n* tapes can be simulated by a Turing machine with one tape [3].

In interaction, machines can change data processed by one of the machines, its software (program of computation) and/or hardware. As Turing machines are utilized as the model of recursive algorithms, hardware is not changed. At the same time, the schema of a universal Turing machine allows one to keep software (rules of computation) in memory in the form of processed data. Consequently, we can assume that only data processed by machines are changed in interaction. That is, interaction goes through memory of machines by changing symbols in memory cells.

## 2   Turing machines with infinite output and interaction

Considering interactive systems and processes, it is necessary to treat concurrent systems and processes. At the same time, as Palamidessi and Valencia, write [8], infinite behavior is ubiquitous in concurrent systems (e.g., browsers, search engines, reservation systems). Thus, it becomes crucial to study and compare systems with tentatively infinite input and output.

Here we can ask a question how it is possible to compare Turing machines that work and were designed to work with finite words and system that process potentially infinite words. If we consider this question in a primitive way, can come to two opposite but mutually simple conclusions. The first one says that if Turing machines cannot work with infinite words and interactive systems can, then interactive systems are evidently more powerful than Turing machines. The second conclusion is that Turing machines and interactive systems are incomparable.

However, the second conclusion brings us to a paradoxical result. Indeed, in the theory of computation and automata, it is proved that finite automata (finite state machines) are weaker than Turing machines. Nevertheless, there are finite automata (for instance, Büchi and Muller automata [2, 11]) that work with infinite words.

There is a simple solution to this paradox. Namely, to show that actually Turing machines can work with infinite words. Turing machines can do this in the same way as finite automata do. Specifically, if an infinite sequence of symbols is given to a Turing machine, the machine can transform it into an infinite output sequence of symbols, separating the input into finite parts and working with one part at a time. However, this transformation of infinite strings of symbols always has a specific property.

Let us assume that all words in the alphabet of considered Turing machines are ordered and all words are given to the considered Turing machine one by one in this order. Then we have the following result.

**Theorem 1.** The output of a Turing machine with infinite output (TMIO) is a recursively enumerable set of finite words and any recursively enumerable set of finite words is an output of a Turing machine with infinite output.

This result remains true even if the input words go in an arbitrary but recursively enumerable order.

**Corollary 1.** If the input of a Turing machine is a recursively enumerable set of finite words, then the corresponding output is also a recursively enumerable set of finite words.

Thus, recursive enumerability is the essence of the Turing machine output even when this output is infinite and input is recursive. More exactly, we have the following result.

**Corollary 2.** a) If the infinite input string is recursively partitioned (divided) into a set of finite words, then the corresponding output of any Turing machine with this input is also a recursively enumerable set of finite words.

b) If the infinite input string is partitioned (divided) into a set of finite words that is not recursively enumerable, then there is a Turing machine such that working with this input, it will as output a set of finite words that is not be a recursively enumerable.

For convenience, we give here a general structure (schema) of a Turing machine $T$ with infinite input/output and interaction:

$$T = (A, R, Q, P, F, q_0, L_I, L_w, L_{iO}, i = 1, 2, 3, \ldots)$$

Here

$A$ is the alphabet of $T$;

$R$ is the system of rules of $T$;

$Q$ is the set of states of $T$;

$P$ is the set of output states of $T$;

$F$ is the set of final states of $T$;

$q_0$ is the start of state of $T$;

$L_I$ is the input tape of $T$;

$L_w$ is the working tape of $T$;

$\{ L_{iO}, i = 1, 2, 3, \ldots \}$ is the system of output tapes of $T$.

It is necessary to remark that a machine can perform infinite computations but have finite input and finite output.

## 3 Types of interactive computing systems and processes

Treating here interactive processes in systems of computational devices (automata or algorithms), we consider their activity as information processing in a general case and as computation when we are interested in their computing power.

Time is important parameter of interactive systems in general and computing systems, which usually consist of various interacting devices, in particular. The most popular model is the linear physical time. However, now some physical theories are based on a two-dimensional model of time [22, 25], is used in the theory of databases [27], and branching time plays an important role in computational models [23, 24].

We consider here situations when interactive systems (processes) interact only in form of information exchange, that is, they communicate. In addition, we assume that interaction goes in a communication space, which is a special media designed for communication [8]. In our theoretical model, we use such a communication space as a linear tape of cells, in which one cell can contain one symbol.

Several interactive processes can go on even in one computing device (information processing system with one processor) when different programs realize these processes. Moreover, even one sufficiently complex program can organize many processes. Operating system of a computer is an example of such a program.

It is possible to divide all interactive computations into three types [5]: free interactive computations, partially free interactive computations, and algorithmic or procedural interactive computations.

In turn, there are two types of procedural (algorithmic) concurrent computations: implicitly procedural (algorithmic) concurrent computations and explicitly procedural (algorithmic) concurrent computations.

**Definition 1.** An interactive computation is called *free* if interactions between processes go without any rules.

For instance, it is demonstrated in [2] that a system of two finite automata interacting without any rules can eventually compute any function. However, when interaction of processes is not specified, at least, by some rules, the enveloping (computational) process can lead to deadlocks, data corruption when different processes change common data without concordance, and other safety violations.

An opposite situation is when all interactions are controlled by definite rules. These rules can be local and global.

**Definition 2.** An interactive computation (functioning of a grid array/automaton) is called *implicitly procedural* (*algorithmic*) if all interactions between processes go according to local rules (where each set of local rules form an algorithm of local interactions).

For instance, each process (algorithm or device) in a system has its own interaction rules. However, for some processes these rules can coincide.

A more rigid type is explicitly procedural algorithmic computation.

**Definition 3.** An interactive computation (functioning of a grid array/automaton) is called *explicitly procedural* (*algorithmic*) if all interactions between processes go according to some system of rules (algorithm).

Algorithmic control of interacting processes is naturally considered as an operation with these processes [5].

**Definition 4.** Explicitly algorithmic functioning of a grid array/automaton is called *algorithmic operation* (AO).

An intermediate situation between free and algorithmic computations is partially free functioning.

**Definition 5.** An interactive computation (functioning of a grid array/automaton) is called *partially free* if not all interactions between processes are specified by rules.

## 4 Sources of interactive superrecursiveness

In this section, we describe five main sources of interactive superrecursiveness. For completeness, we present here results describing the first three evident sources. The first one is when the algorithms itself is super-recursive. The second is also simple and is described in the following proposition.

**Proposition 1.** An interactive recursive algorithm (a Turing machine) *A* can compute a recursively non-enumerable set if it can contain a recursively non-enumerable set as its initial information.

Indeed, taking a recursively non-enumerable set that constitutes initial information, the machine *A* gives this set as its output. To do this, we even do not need such a powerful model as a Turing machine. In this case, superrecursive computation can be realized by a finite automaton *A* that is provided from the start with a recursively non-enumerable set as its input.

Another trivial cause for interactive superrecursiveness is interaction with a system that can send to algorithm (automaton) *A* recursive incomputable information. An example of such a system is a non-recursive oracle, with which a Turing machine can interact [10]. However, we are interested in interaction of algorithms (automata). Consequently, we consider the second system as a super-recursive algorithm.

**Proposition 2.** An interactive recursive algorithm (a Turing machine) *A* can compute a recursively non-enumerable set if it interacts with a super-recursive algorithm *B*.

Indeed, the algorithm *B* can compute a recursively non-enumerable set and give it to the machine *A*. Then the machine *A* acts as in the previous case. Namely, receiving a recursively non-enumerable input, the machine *A* gives it as its output. To do this, we even do not need such a powerful model as a Turing machine. In this case, superrecursive mode of functioning can be also realized by a finite automaton *A*.

One more source of interactive superrecursiveness is time. This is possible when interaction is not synchronized either because automata (algorithms) send and receive information at random (or at least, at non-recursively coordinated) moments of time or due to incommensurability of the time scales in which automata work.

**Theorem 2.** An interactive recursive algorithm (a Turing machine) *A* can compute a recursively non-enumerable set if the temporal sequence of information exchanges is recursively non-enumerable.

This result is proved in [2] for the case when interaction of system automata with the communication space is random and thus, non-enumerable [3].

It is necessary to remark that machines with random interaction do not compute a function on infinite words (strings). The result of their computation is a multivalued function. However, it possible to have a deterministic computation on infinite words and still to achieve super-recursive results. To do this, it is necessary to have a source that controls interaction of system automata with the communication space in a super-recursive way, e.g., the sequence of moments when machines have access to this space is non-enumerable.

**Theorem 3.** An interactive recursive algorithm (a Turing machine) $A$ can compute a recursively non-enumerable set when the time scales in which automata work are not commensurable.

What does it mean incommensurability of the time scales of two automata $A$ and $B$? In a general case, it means that while the automaton $A$ makes some number of moves (say, $n$), the automaton $B$ can make any number of moves (say, $m$).

Now let us consider two automata $A$ and $B$. The automaton $A$ writes the symbol 0 into the common output tape every other move of its functioning, while each odd move the automaton $B$ includes writing the symbol 1 into the common output tape. Symbols are written from left to right, and each time the first free cell to the right is filled. When these automata work synchronously, their output will have the form 10101010 … However, when the time scales in which automata work are not commensurable, it is possible that while the automaton $B$ makes one move, the automaton $A$ makes ten moves. After this, both automata work synchronously. In this case, their output will have the form 0000010101010 … If such a situation can happen not only at the beginning, but also at any stage of computation, the output of $A$ and $B$ can be not recursively enumerable.

In this case, incommensurability is not uniform. That is, one time when the automaton $A$ makes $n$ moves, the automaton $B$ makes $m$ moves, but another time when the automaton $A$ makes $n$ moves, the automaton $B$ makes $k$ moves. Formally it means that intervals with the same length in the first time scale can be mapped to intervals with different lengths in the second time scale.

Thus, it is interesting to find whether superrecursiveness is possible when incommensurability of the time scales is uniform. Formally it means that intervals with the same length in the first time scale are always mapped to intervals with equal lengths in the second time scale. When we consider physical time, the time scales are isomorphic to the real line. In this case, incommensurability of the scales means that the time unit of one scale is mapped to some irrational interval in the second scale. Mathematically such a mapping is a stretching or contraction with an irrational parameter.

**Theorem 4.** An interactive recursive algorithm (a Turing machine) $A$ can compute a recursively non-enumerable set when the time scales in which automata work are uniformly incommensurable.

One more source of interactive superrecursiveness is space. This is possible when do not determine where they put their data in the communication space.

**Theorem 5.** An interactive recursive algorithm (a Turing machine) $A$ can compute a recursively non-enumerable set if the sequence of cells where data written by another system is recursively non-enumerable.

To prove this we consider two Turing machines $A$ and $B$ that interact through a communication space. This space is a linear one-sided tape. While functioning, the machines $A$ and $B$ write symbols 1 or 0 in cells of the interaction tape.

The machine $B$ works in a very simple way. At the step $n$, it puts the symbol 1 into one of two cells with numbers $2n - 1$ and $2n$. The other cell is filled by $A$ with the symbol 0. By the assumption of the theorem, it is possible that the sequence of zeros and ones is recursively non-enumerable.

After the pair of cells with numbers $2n - 1$ and $2n$ is filled, the machine $A$ gives the output 0 if it was 01 in those two cells and the output 1 if it was 10 in those two cells. When the sequence of zeros and ones in the communication space is recursively non-enumerable, then the output sequence of the machine $A$ is also recursively non-enumerable.

## 5  Interaction with a recursive control

There are two main kinds of algorithmic control of interaction: global and local. In global control, there is an algorithm (automaton) $C$ that controls interaction of $A$ and $B$. Namely, $C$ determines when and where $A$ and $B$ read from and write into the communication tape. In addition, time scales $L_A$ and $L_B$ of both $A$ and $B$ are synchronized with time scale $L_C$ of $C$. That is, temporal units $1_A$ and $1_B$ (e.g., second, millisecond, etc.) of each scale $L_A$ and $L_B$ are equal to $n1_C$ and $m1_C$ for some whole numbers $n1_A$ and $m1_B$ where $1_C$ is a temporal unit of the scale $L_C$.

In local control, each machine $A$ and $B$ determines when and where $A$ and $B$ read from and write into the communication tape. An additional machine $C$ only solves conflicts when both $A$ and $B$ try to write different symbols into the same cell. In addition, all three time scales $L_C$, $L_A$ and $L_B$ are synchronized.

Theorem 4 shows that local rules are insufficient to restrict the computing power of the interacting recursive devices to recursive computations. Only a global algorithm (device) that organizes interaction can do this.

**Theorem 6.** An interactive system (grid automaton) $R$ that consists of a finite number of recursive devices (algorithms) interaction of which is organized by a recursive automaton (algorithm) can perform only recursive computations (may be, infinite), i.e., such a system $R$ is equivalent to a Turing machine.

Note that in this case time scales of all automata from $R$ are synchronized with the time scale of the control automaton and thus, they are synchronized with one another.

## 6  Conclusion

Thus, we have found three trivial (the system itself, initial information or/and another system is super-recursive) and two (or actually, three because the temporal parameter provides two possibilities) non-trivial (time and space) causes for interactive superrecursiveness. In addition, we demonstrate that this list of causes for interactive superrecursiveness is complete.

It would be interesting to study other situations when a system of interacting devices (automata or processes) can have higher computational power than the power of each constituent of this system. Here we did this for systems consisting of Turing machines as it is the most popular model of computation and as it caused the most active controversy. However, the same problem can be studied for subrecursive models, such as finite and pushdown automata, and super-recursive models, such as inductive and limit Turing machines [4].

There are also important problems with analyzing the concept of computation:

What is computation?

Is computation always algorithmic?

Is computation always a physical process?